\documentstyle[12pt]{article}

\newcommand{\sect}[1]{\setcounter{equation}{0}\section{#1}}

\def\bseq{\begin{subequation}}  
\def\eseq{\end{subequation}}
\def\bsea{\begin{subeqnarray}}  
\def\esea{\end{subeqnarray}}

\def\beq{\begin{equation}}
\def\eeq{\end{equation}}
\def\eea{\end{eqnarray}}
\def\bq{\begin{quote}}
\def\eq{\end{quote}}

\newcommand{\EQ}{\begin{equation}}
\newcommand{\EN}{\end{equation}}
\newcommand{\bea}{\begin{eqnarray}}
\newcommand{\ena}{\end{eqnarray}}

\renewcommand{\a}{\alpha}
\renewcommand{\b}{\beta}

\renewcommand{\d}{\delta}

\newcommand{\pa}{\partial}
\newcommand{\g}{\gamma}

\newcommand{\e}{\epsilon}

\newcommand{\h}{\eta}

\renewcommand{\l}{\lambda}
\renewcommand{\L}{\Lambda}
\newcommand{\m}{\mu}

\newcommand{\n}{\nu}

\newcommand{\r}{\rho}

\newcommand{\s}{\sigma}

\renewcommand{\o}{\omega}

\newcommand{\del}{\partial}

\renewcommand{\thefootnote}{\fnsymbol{footnote}}

\begin{document}
\newpage
\begin{titlepage}
\begin{flushright}
{CERN-TH.6458/92\\
hepth@xxx/9205093}
\end{flushright}
\vspace{2cm}
\begin{center}
{\bf {\large Induced (N,0) supergravity as a constrained Osp(N$|$2)}}\\
\vspace{.1in}
{\bf {\large  WZWN model and its effective action}} \\
\vspace{1.5cm}
G.W. DELIUS,
M.T. GRISARU\footnote{On leave from Brandeis University, Waltham, MA 02254,
USA\\Work partially supported by the National Science Foundation under
grant PHY-88-18853.} \\
\vspace{3mm}
and\\
\vspace{3mm}
P. VAN NIEUWENHUIZEN\footnote{On leave from the Institute for Theoretical
Physics of the State University of New York at Stony Brook, Stony Brook,
N.Y. 11794, USA. Work partially supported by the National Science Foundation
under grant PHY-89-08495.}\\
\vspace{1mm}
{\em Theory Division, CERN, 1211 Geneva 23, Switzerland}\\
\vspace{5mm}

\vspace{1.1cm}

{\bf ABSTRACT}
\end{center}
\bq
A chiral $(N,0) $ supergravity theory in d=2 dimensions for any $N$ and its
induced action can be obtained
by constraining the currents of an Osp(N$|$2) WZWN
model. The underlying symmetry algebras are the nonlinear SO(N) superconformal
algebras of Knizhnik and Bershadsky. The case $N=3$ is worked out in detail.
We show that by
adding quantum corrections to the classical transformation rules, the gauge
algebra on gauge fields and currents closes. Integrability conditions on
Ward identities are derived. The effective action is
computed at one loop. It is finite, and can be obtained from the induced
action by
rescaling the central charge and fields by finite
Z factors.

\eq

\vfill

\begin{flushleft}
CERN-TH.6458/92\\
May 1992
\end{flushleft}
\end{titlepage}

\renewcommand{\thefootnote}{\arabic{footnote}}
\setcounter{footnote}{0}
\newpage
\pagenumbering{arabic}
\sect{Introduction.}

Recently, it has been conjectured, and verified at the one-loop level, that
gauge theories in two-dimensional Euclidean space
have remarkable renormalizability properties. Namely, their full effective
actions are free from any divergences, and
obtained from their induced actions merely by rescaling the
coupling constants and the fields by constant $Z$-factors, and these
$Z$-factors are power series in the inverse of the coupling constants.
The induced actions can be obtained by first coupling matter to external gauge
fields in a minimal way, and then integrating out the matter. In other words,
the induced actions are the set of 1PI diagrams with propagating matter and
external gauge fields. Since the gauge symmetry of these classical actions is
broken at the quantum level by anomalies, a quadratic term and vertices
involving
these fields develop, so that loops with propagating gauge fields
can be constructed. The induced actions are nonlocal.

Both the induced actions and the effective actions are finite (i.e., no
divergences are generated by the loops) and both depend only on the gauge
fields, but whereas the former are due to diagrams with matter loops and
external gauge fields, the latter are due to diagrams with gauge fields
both in the loops and as external fields.
In local quantum field
theories loops produce nonlocalities, needed for unitarity, and there is
no way that the complete nonlocal effective action can be obtained from
the local quantum action by constant rescalings. However, in the class
of theories we are going to consider, the nonlocalities due to gauge loops
are exactly of the same form as the nonlocalities present in the induced
actions, and this allows the effective actions to be obtained from the
induced actions in the manner described above.

To date, three models are known with these properties:

\noindent (i) Yang-Mills theory, whose induced action is the WZWN model, but
in a purely two-dimensional nonlocal formulation [1].

\noindent (ii) ordinary gravity, whose induced action is Polyakov's chiral
gravity [1,2].

\noindent (iii) $W_3$ gravity, based on the nonlinear $W_3$ algebra of
Zamolodchikov. Its induced and (one-loop) effective actions were
 obtained in refs. [3,4,5] and contain
a spin
2 gauge field $h$ and a spin 3 gauge field $b$.

To this list we shall add a fourth example: gauge theories based on the
nonlinear SO(N) algebras of Knizhnik [6] and Bershadsky [7].
The interest of these models is that they lead to chiral supergravities
for any N. They lie between the models in (ii) and
(iii): they are based on a nonlinear algebra (as $W_3$ gravity) but contain no
higher spins (as Polyakov's gravity). By a nonlinear algebra we mean a set
of operators whose operator product expansion (OPE) contains terms quadratic
(or higher) in the operators. In the $W_3$ algebra, one finds in the OPE of
two spin 3 currents two terms bilinear in the stress tensor. In the SO(N)
models one finds in the OPE of two spin 3/2 (supersymmetry) currents a term
bilinear in the SO(N) Ka\v{c}-Moody currents. The corresponding induced actions
describe the gauge fields $h$ (graviton), $\psi^i$ (gravitini), and $\o^a$
( SO(N) gauge vector fields).

It may be useful to be a bit more explicit about these models. Therefore we
briefly summarize the main results for the first three models.

{\bf (i) Yang-Mills theory.} If one couples a chiral gauge field
$A_+(x^+,x^-) = A_+^aT_a$ (with $[T_a,T_b] =f_{ab}^{~~c}T_c$) to chiral
matter fermions, and integrates over these fermions (i.e., evaluates all
one-loop diagrams with internal fermions and any number of external $A_+$
fields), one finds the following result for the induced action [3]
\footnote{Conventions for (super)algebras:
$(-1)^p f_{pa}^{~~q}f_{qb}^{~~p} = -g_{ab} \tilde{h}$,
$str T_aT_b = -\chi g_{ab}$. Raising and lowering is done with $g_{ab}$ and its
inverse $g^{ab}$; $\tilde{h}$ is the dual Coxeter number.}
\bea
S_{ind}[k,A_+] &=& k S_{ind}^{(0)} [A_+] \nonumber\\
S_{ind}^{(0)} &=& - \frac{1}{2\pi \chi} \sum_{n=0}^{\infty} \frac{1}{2+n}
tr \int d^2x A_+ \left[ \frac{1}{\pa_+}A_+ , \cdots \left[\frac{1}{\pa_+}A_+ ,
\frac{\pa_-}{\pa_+}A_+ \right] \cdots \right]_{n~times}
\nonumber\\
&=& \frac{1}{2\pi} \int d^2x \left[ \frac{1}{2} A_+^a \frac{\pa_-}{\pa_+}A_{+a}
+\frac{1}{3}f_{abc}A_+^a \frac{1}{\pa_+}A_+^b \frac{\pa_-}{\pa_+}A_+^c +
\cdots \right]
\ena

Under $\d A_+^a = \pa_+\eta^a -f^a_{~bc}A_+^b\eta^c$ the induced action
varies into $\frac{k}{2\pi} \int d^2x A_+^a\pa_-\eta_a$. Hence, writing $\d
S = (\d S/\d A)\d A$, it  satisfies the
following Ward identity
\EQ
\pa_+u_-^a - \pa_-A_+^a - f^a_{~bc}A_+^bu_-^c =0
\EN
where $u_-^a$ are the currents, suitably normalized,
\EQ
u_{-,a} = \frac{2\pi}{k} \frac{\d S_{ind}}{\d A_+^a} = \frac{\pa_-}{\pa_+}
A_{+,a} + \cdots
\EN

In fact, any matter system with Ka\v{c}-Moody current $J_a(x)$ satisfying
the OPE
\EQ
\underbrace{J_a(x)J_b(y)}
= -\frac{k}{2}\frac{g_{ab}}{(x^--y^-)^2} +\frac{f_{ab}^{~~c}
J_c(y)}{x^--y^-}
\EN
leads to this induced action if one adds to the matter action the
following minimal coupling
\EQ
S_{int} =\frac{1}{\pi} \int d^2x A_+^a(x)J_a(x)
\EN
For example, a WZWN model itself is such a matter system, although in
this case the induced action receives contributions from
arbitrarily many matter loops.

The {\em effective}  action, i.e., the sum of all 1PI graphs computed from
(1.1) with internal and
external $A_+$ lines, is conjectured [1,8] to be given by
\EQ
S_{eff}[A] =Z_k k S_{ind}^{(0)} [Z_A A]
\EN
where the $Z$ factors are constant, and power series in the inverse of
the level parameter $k$ of the Ka\v{c}-Moody current [8] \footnote{In ref. [1]
the result for $Z_A$ differs from our result by a factor of 2 in
front of $\tilde{h}$.}
\EQ
Z_k = 1+\frac{2\tilde{h}}{k} ~~~~;~~~~
Z_A = \left(1+\frac{\tilde{h}}{k} \right)^{-1}
\EN
and $\tilde{h}$ is the dual Coxeter number. To one loop this conjecture
has been verified (see below).

{\bf (ii) Polyakov gravity.} If one couples the component $h \equiv
h_{++}$ of the gravitational field to the stress tensor of matter, and
integrates over matter, one finds the action studied by Polyakov,
Zamolodchikov, and Knizhnik [2]
\EQ
S_{ind} [c,h] = c S_{ind}^{(0)}[h]
\EN
with
\bea
S_{ind}^{(0)} [h] &=& -\frac{1}{24\pi} \int d^2x( \pa_-^2 h) \frac{1}
{1- \frac{1}{\pa_+}h\pa_-} \frac{\pa_-}{\pa_+} h \\
&=& -\frac{1}{24\pi} \int d^2x \left[ h\frac{\pa_-^3}{\pa_+}h -
h\left(\frac{\pa_-^2}{\pa_+}h\right)^2 -\left(h \frac{\pa_-^2}{\pa_+}h\right)
\frac{\pa_-}{\pa_+}\left(h \frac{\pa_-^2}{\pa_+}h \right) + \cdots \right]
\nonumber
\ena
This result holds whenever the OPE for $T \equiv T_{--} $ has the
standard form
\EQ
\underbrace{T(x)T(y)} = \frac{c}{2} \frac{1}{(x^--y^-)^4} +2
 \frac{T(y)}{(x^--y^-)^2}
+ \frac{T'(y)}{x^--y^-}
\EN
The effective action has been conjectured [9] to be given by
\EQ
S_{eff}[c,h] = Z_c c S_{ind}^{(0)}[Z_h h]
\EN
where
\bea
Z_c &=& \frac{6}{c}k_c = 1-\frac{25}{c} +\cdots
{}~~~~;~~~~~Z_h=\frac{k_c+2}{k_c} =1+\frac{12}{c} +\cdots\nonumber\\
c-13 &=&6[(k_c+2) +(k_c+2)^{-1}] \nonumber\\
k_c+2 &=&-\frac{1}{12}[13-c +\sqrt{(c-1)(c-25)} ] ~~~~{\rm for}~ c<0
\ena

{\bf (iii) $W_3$ gravity.} Coupling the chiral gauge fields $h \equiv h_{++}$
with spin 2 and $b \equiv b_{+++}$ with spin 3 to matter
currents $T_{--}$ and $W_{---}$ which satisfy the exact
$W_3$ algebra of Zamolodchikov with central charge $c$, one finds that
the induced action is a power series in $1/c$ [4]
\bea
S_{ind}[c,h,b] &=& \sum_{j=0}^{\infty} c^{1-j} S_{ind}^{(j)}[h,b]
 \nonumber\\
S_{ind}^{(0)}[h,b] &=& -\frac{1}{24\pi} \int d^2x h \frac{\pa_-^3}{\pa_+}h
-\frac{1}{720\pi} \int d^2x b\frac{\pa_-^5}{\pa_+}b +\cdots
\ena
The effective action, due to $h$ and $b$ loops with vertices from all
$S^{(j)}_{ind}$, is conjectured to be obtained by just
rescaling $cS_{ind}^{(0)}$
[5]
\EQ
 S_{eff}[c,h,b] = Z_c c S_{ind}^{(0)}[Z_hh, Z_bb]
\EN
where
\bea
&& Z_c=\frac{24}{c}k_c = 1-\frac{122}{c}+\cdots ;
{}~~~~;~~~    ~ Z_h=\frac{k_c+3}{k_c}=1+\frac{72}{c}+\cdots  \nonumber\\
&&  Z_b =\frac{(k_c+3)^{3/2} \sqrt{30\b}}{2k_c}
 = 1+\frac{224}{5c}+\cdots \nonumber\\
&& c-50 = 24\left[(k_c+3) +(k_c+3)^{-1} \right] \nonumber\\
&& k_c+3 = -\frac{1}{48}\left[ 50-c +\sqrt{(c-2)(c-98)}\right] ~~~{\rm for}~
c<0
\ena
Here $\b = 16(22+5c)^{-1}$ is a constant which appears in the OPE of
$W_{---}(x)$ and $W_{---}(y)$ in front of the terms with two $T_{--}$
operators.

Before closing this introduction we review how one computes the one-loop
contribution to the effective action since by this method we shall determine
the $Z$-factors.  We shall do this for
Yang-Mills theory.

For Yang-Mills theory, one starts from the  Ward identity in eq. (1.2).
(It states that $u_- = \pa_-g g^{-1}$ and $A_+= \pa_+g g^{-1}$ for some $g$,
and if
one were to make this substitution for $A_+$ in $S_{ind}[A_+]$, one
would recover the WZWN model in a form where the usual local
3-dimensional WZWN term is written as a nonlocal 2-dimensional term.
However, we do not make this substitution and keep working with $A_+$,
because only in terms of $A_+$ do we have the remarkable renormalizability.)
By differentiating the Ward identity one finds a relation of the form
\EQ
M^a_{~b}(x)\left( \frac{\pa u^b_-(x)}{\pa A_+^c(y)}\right) = N^a_{~c}(x)\d
(x-y)
\EN
with
\bea
M^a_{~b} &=& \d ^a_{~b} \pa_+ + f^a_{~bc}A_+^c  \nonumber\\
N^a_{~b} &=& \d^a_{~b}\pa_- + f^a_{~bc}u_-^c
\ena
Since the one-loop contributions are given by the determinant of the matrix
$\pa u(x)/ \pa A(y)$, it
suffices to evaluate $\det M$ and $\det N$. Both $M$ and $N$ are local
matrix-operators, and we obtain their determinants by using ``ghosts''
$b_a$, $c^a$ for $M$, and $B_a$, $C^a$ for $N$, i.e.,
\bea
\det M &=& \int dc ~db \exp[b_a M^a_{~b}c^b] \nonumber\\
\det N &=& \int dC~dB \exp[ B_aN^a_{~b}C^b ]
\ena

To actually evaluate these determinants, one introduces the notation
\EQ
\det
M = \left\langle \exp \int b_aj^a_{~b}c^b \right
 \rangle ~~~~,~~~j^a_{~b}=f^a_{~bc}A_+^c
\EN
where the expectation value is taken with respect to the free ghost actions,
and uses
\EQ
\langle c^a(x) b_b(y) \rangle = - \d ^a_{~b} \frac{1}{\pa_+}\d^2(x-y)
\EN
One finds then
\EQ
\det M = 1- \frac{1}{2\pi} tr \int j \frac{\pa_-}{\pa_+}j +
\frac{1}{3\pi} tr \int j \left[\frac{1}{\pa_+}j , \frac{\pa_-}{\pa_+}j
\right] +\cdots
\EN
Evaluating the traces one finds
\EQ
\frac{1}{2} \ln \det M = \tilde{h} \left[ \frac{1}{4\pi} \int A_+^a
\frac{\pa_-}
{\pa_+} A_{+a} + \frac{1}{6\pi}f_{abc}\int A_+^a \left( \frac{1}{\pa_+}A_{+}^b
\right) \left(\frac{\pa_-}{\pa_+}A_+^c\right) +\cdots \right]
\EN
In other words, $\frac{1}{2} \ln \det M$ is proportional to the induced action
\EQ
\frac{1}{2}\ln \det M =\tilde{h} S_{ind}^{(0)} [A_+]
\EN
To prove this to all orders in $A_+$, one may use the Ward identity for
(1.18) [1,8].

For $N$ one finds a slightly different result. Of course $\det N$ is the
same functional as $\det M$, but with $j^a_{~b}$ replaced by
$J^a_{~b}=f^a_{~bc}u_-^c$, and with $\pa_-$ and $\pa_+$ interchanged. Hence
\EQ
\frac{1}{2} \ln \det N =\tilde{h} \bar{S}_{ind}^{(0)} [u_-]
\EN
where we denote by $\bar{S}_{ind}^{(0)}[u_-]$ the induced action with
$A_+$ replaced by $u_-$ and $\pa_+$ and $\pa_-$ interchanged.
(Also this result holds to all
orders as can be shown [1,8] by using the Ward identity for $\det N$ in
(1.18).)
However, if we substitute for $u_-^a$ its dependence on $A_+$
\EQ
u_-^a = \frac{\pa_-}{\pa_+}A_+^a +\frac{1}{3}f^a_{~bc} \left\{
\left(\frac{1}{\pa_+}A_+^b\right)\left( \frac{\pa_-}{\pa_+}
A_+^c \right) +\frac{1}{\pa_+}\left(A_+^b\frac{\pa_-}{\pa_+}
A_+^c \right)   +\frac{\pa_-}{\pa_+}\left( A_+^b \frac{1}{\pa_+}A_+^c\right)
\right\} +\cdots
\EN
we find
\EQ
\frac{1}{2} \ln \det N = \tilde{h} \left[ \frac{1}{4\pi} \int A_+^a
\frac{\pa_-}{\pa_+} A_+^a +\frac{1}{3\pi}f_{abc}\int
A_+^a\left(\frac{1}{\pa_+}A_+^b \right) \left(
\frac{\pa_-}{\pa_+}A_+^c\right) +\cdots \right]
\EN

To this order at least, $\bar{S}_{ind}^{(0)}[u_-(A_+)]$ is related
to $S_{ind}^{(0)}[A_+]$
by a Legendre transform [1]
\EQ
S_{ind}^{(0)}[A_+] +\bar{S}_{ind}^{(0)}[u_-(A_+)] = \frac{1}{2\pi}
\int d^2x A_+^a u_{-a}
\EN
This is easy to check; $Au \sim A \d S_{ind} /\d A$ counts the number of
$A$-fields in $S_{ind}$, and multiplies the terms with 2,3,... $A_+$
fields in $S_{ind}^{(0)}$ by 2,3,.. .
Therefore the complete one-loop contribution to the
effective action is given by
\EQ
S_{eff}^{1-loop} =\frac{1}{2} \ln \det M - \frac{1}{2} \ln \det N =
2\tilde{h} S_{ind}^{(0)}- \frac{\tilde{h}}{2\pi} \int d^2x A_+^a u_{-a}
\EN
In other words, the one-loop contributions replace $k$ in front of the
induced action by $2\tilde{h}$ and scale each field $A_+$ with a factor
$-\tilde{h}$.
This implies
\EQ
S_{ind}[A_+] +S_{eff}^{1-loop}[A_+] = (k+2\tilde{h})S_{ind}^{(0)}\left[(1-
\frac{\tilde{h}}{k})A_+ \right]
\EN
The all-loop result is conjectured [8] to be obtained by replacing
$1-\tilde{h}/k$ by $(1+\tilde{h}/k)^{-1}$.

\sect{The SO(N) extended superconformal algebras.}

These algebras [6,7]
contain a current of dimension 2 (the stress tensor $T(z)$),
$N$ currents of dimension 3/2 (the supersymmetry currents $Q^i(z)$, with
$i=1,...N$), and $\frac{1}{2}N(N-1)$ currents of dimension 1 (the
Ka\v{c}-Moody currents $J^a(z)$ for SO(N)). We shall concentrate on SO(3),
because it is the simplest interesting case. (For N=2 the algebra
becomes linear.) The OPE for these currents is as expected:
$\underbrace{T(z)T(w)}$
was given in eq. (1.10), and $Q^i$, $J^a$ are primary:
\bea
\underbrace{T(z)Q^i(w)} &=& \frac{3}{2} \frac{Q^i(w)}{(z^--w^-)^2}
 +\frac{Q'^i(w)}{z^--w^-}
\nonumber\\
\underbrace{T(z)J^a(w)} &=& \frac{J^a(w)}{(z^--w^-)^2} +\frac{J'^a(w)}{z^--w^-}
\ena
Furthermore, the Ka\v{c}-Moody currents act on themselves and on $Q^i(z)$ as
SO(3) generators
\bea
\underbrace{J^a(z)J^b(w)} &=&- \s \frac{\d^{ab}}{(z^--w^-)^2}+\epsilon^{abc}
 \frac{J_c(w)}
{z^--w^-} \nonumber\\
\underbrace{J^a(z)Q^i(w)} &=&  \epsilon^{aij} \frac{Q_j(w)}{z^--w^-}
\ena
However, the $QQ$ OPE contains a nonlinear term
\bea
\underbrace{Q^i(z)Q^j(w)} &=& B \frac{\d^{ij}}{(z^--w^-)^3} -K \epsilon^{ija}
 \frac
{J_a(w)}{(z^--w^-)^2} -\frac{K}{2} \epsilon^{ija}\frac{J'_a(w)}{z^--w^-}
\nonumber\\
&+& \d^{ij} \frac{2T(w)}{z^--w^-} +2\g\frac{ J^{ij}(w)}{z^--w^-}
\ena
where $J_{ij}$ is the normal-ordered product of two Ka\v{c}-Moody currents,
symmetrized in the indices
\bea
J_{ij} &=& \frac{1}{2} :J_iJ_j: +\frac{1}{2}:J_jJ_i: ~\equiv ~
:J_{(i}J_{j)}: \nonumber\\
:J_iJ_j:(w) &=& \frac{1}{2\pi i} \oint\frac{dx}{x-w}J_i(x)J_j(w)
\ena
There is only one independent central charge, which we choose to be $\s$,
as the Jacobi identities
require the following relations
\bea
c&=& \frac{1}{2}(6\s -1) ~~~~,~~~K=\frac{2\s -1}{\s} \nonumber\\
B&=&K\s =2\s -1 ~~~~,~~~\g = \frac{1}{2\s}
\ena
To check these results one may compute, for example,
\bea
\langle J_a(z) Q_i(x)Q_j(y)\rangle &=& \langle \underbrace{J_a(z)Q_i(x)}
Q_j(y) \rangle +\langle Q_i(x) \underbrace{J_a(z)Q_j(y)} \rangle
\nonumber\\
&=& \langle \underbrace{Q_i(x)J_a(z)}Q_j(y) \rangle +\langle J_a(z)
\underbrace{Q_i(x)Q_j(y)}\rangle
\ena
The first way of contracting yields
\EQ
B\e_{aij} \frac{1}{(x^--y^-)^3}
\left(\frac{1}{z^--x^-}-\frac{1}{z^--y^-}\right)
\EN
while the second way yields
\EQ
B\e_{aij}\frac{1}{(z^--x^-)(z^--y^-)^3} +\s K\e_{aij}
\left( \frac{1}{(x^--y^-)^2(z^--y^-)^2}
+\frac{1}{(x^--y^-)(z^--y^-)^3} \right)
\EN
Clearly $\s K=B$. \footnote{In ref. [6], a factor $1/2$ is missing in
the $QQ \sim J$ term, while in ref. [7] this term has an incorrect sign.}

Decomposing the currents in modes, the SO(N) algebras are of the general form
\bea
{~}[ H_i,H_j ] &=& f_{ij}^{~~k}H_k +h_{ij}I \nonumber\\
{~}[ H_i,S_{\a} ] &=&f_{i\a}^{~~\b}S_{\b} +f_{i\a}^{~~j}H_j \nonumber\\
{~}[S_{\a},S_{\b}] &=&f_{\a \b}^{~~~i}H_i +f_{\a \b}^{~~~\g}S_{\g}
+V_{\a \b}^{~~~ij} :H_iH_j: +h_{\a \b}I
\ena
where we can, without loss of generality, take the constants $V_{\a
\b}^{~~~ij}$
to be symmetric in $ij$. Indeed, if $H_i = \{J_a\}$ and $S_{\a} = \{T, Q_i\}$,
one obtains this structure. Another division of generators with this
structure is $H_i = \{J_a, T\}$ and $S_{\a}= \{Q_i\}$. For algebras
of this kind, a nilpotent BRST operator exists, provided the central
charges $h_{ab}$ and the structure constants satisfy a relation of the form
[10]
\EQ
h_{ab} \sim F^d_{~ac}F^c_{~db}
\EN
where the index $a$ denotes both $\a$ and $i$. The $F_{ab}^{~~c}$ are
equal to the classical $f_{ab}^{~~c}$ plus, for $f_{\a \b}^{~~j}$, a
correction term of the form $f_{\a i}^{~~\g} V_{\g \b}^{~~ij}$. The
BRST charge reads then [10]
\bea
Q &=& c^aT_a -\frac{1}{2} \bar{c}_cF^c_{~ab}c^ac^b -\frac{1}{2}V_{\a
\b}^ {~~ij} T_i \bar{c}_j c^{\a}c^{\b} \nonumber\\
&-& \frac{1}{24}V_{\a \b}^{~~ij}V_{\g \d}^{~~kl}F^{~~m}_{ik}\bar{c}_j\bar{c}_l
\bar{c}_m c^{\a}c^{\b}c^{\g}c^{\d}
\ena
(In the last term one may clearly replace $F^{~~m}_{ik}$ by $f^{~~m}_{ik}$.)

For the SO(N) algebras the condition (2.10) is satisfied and the
structure constants become multiplicatively renormalized, namely such that in
the structure constants for $QQ \sim J$ the factor $K$ is replaced by
$\frac{1}{2}$,
provided
\EQ
\s =6-2N ~~~,~~~B=16-6N ~~~,~~~ c=N^2-12N+26
\EN
For N=3 (our case), however, one finds that $\s =0$, hence $V_{\a \b}
^{~~~ij}$ which is proportional to the constant $\g = \frac{1}{2}\s^{-1}$,
becomes singular. First multiplying $Q$ for general N by $\g^{-2}$ and then
taking N=3 leaves only the last term in $Q$, which is trivially
nilpotent. At N=3 also $K \rightarrow - \infty$, so that presumably no
unitary irreducible representations exist. In any event, we are considering
general values of $\s$ and these issues are of no concern for us.

\sect{The Ward identities for the SO(3) induced action.}

We define the induced action for the gauge fields $h$, $\psi^i$, $\o^a$, by
\bea
&&e^{S_{ind}[\s ,h,\psi ,\o ]} = \langle e^{S_{int}} \rangle \nonumber\\
&&S_{int} =-\frac{1}{\pi}\int d^2x (hT+\psi^iQ_i +\o^aJ_a)
\ena
Assuming that $\langle  T \rangle =\langle Q_i \rangle = \langle J_a
\rangle =0 $, and expanding the exponential, we can use the OPE given in
the previous sections to determine $S_{ind}$. For example, the kinetic
terms are found to be
\EQ
S_{ind}^{kin} = -\frac{c}{24\pi}\int h \frac{\pa_-^3}{\pa_+} h -
\frac{B}{4\pi}\int\psi^i \frac{\pa_-^2}{\pa_+}\psi_i +\frac{\s}{2\pi}
\int \o^a \frac{\pa_-}{\pa_+}\o_a
\EN
In general, the induced action is an infinite series in inverse
powers of the independent central charge $\s$.
As in the $W_3$ gravity case we write
\EQ
S_{ind}(\s ,h,\psi ,\o )= \sum_{j=0}^{\infty} \s^{1-j}S_{ind}^{(j)}[h,
\psi , \o
]
\EN
The Ward identities for the induced action can be obtained by varying $\exp
[S_{ind}]$ under the leading terms in the variations of the gauge fields
\EQ
\d h = \pa_+\e +\cdots ~~~,~~~ \d \psi^i = \pa_+\eta^i +\cdots ~~~,~~~
\d \o^a = \pa_+\l^a+\cdots
\EN
and then finding extra terms in the transformation laws such that only
the minimal anomalies remain, plus terms due to the nonlinearity of the
algebra. The minimal anomalies, which correspond to the central terms in the
OPE,  are obtained by substituting the above
variations into (3.1) and retaining only the terms quadratic in
operators; hence the result is the same as obtained by
substituting (3.4) directly into
(3.2)
\EQ
{\rm Minimal~ anomaly}
= -\frac{c}{12\pi} \int d^2x h\pa_-^3 \e -\frac{B}{2\pi} \int d^2x \psi^i
\pa_-^2 \eta_i +\frac{S}{\pi} \int d^2x \o^a \pa_-\l_a
\EN

For the local $\e$ symmetry one finds that under $\d h = \pa_+ \e$
\bea
\d S_{ind} \exp [S_{ind}] &=& \langle -\frac{1}{\pi} \int d^2x \pa_+ \e T
e^{S_{int}} \rangle \nonumber\\
&=&\left \langle -\frac{1}{\pi}\int d^2x \pa_+\e
 \underbrace{T
(-\frac{1}{\pi}\int d^2y (hT+\psi^iQ_i +\o^aJ_a}) e^{S_{int}} \right \rangle
\ena
The  central term from $\langle TT \rangle $ yields the minimal
$\e$-anomaly, while all other terms in the OPE are linear in operators,
and are cancelled by suitable extra $\d h$, $\d \psi^i $, and $\d \o^a$.
One finds then that
\EQ
\d(\e )S_{ind} = -\frac{c}{12\pi} \int d^2x h \pa_-^3 \e
\EN
under
\bea
\d (\e )h &=& \pa_+\e -h \pa_-\e +\e \pa_-h \nonumber\\
\d (\e ) \psi^i &=& -\frac{1}{2}\psi^i \pa_-\e +\e \pa_-\psi^i
 \nonumber\\
\d (\e ) \o^a &=& \e \pa_-\o
\ena
Introducing suitably normalized currents by
\bea
u &=& - \frac{12\pi}{c} \frac{\d }{\d h}S_{ind} = \frac{\pa_-^3}{\pa_+}h +
\cdots \nonumber\\
q_i &=& -\frac{2\pi}{B} \frac{\d }{\d \psi^i}S_{ind} = \frac{\pa_-^2}{\pa_+}
\psi_i +\cdots \nonumber\\
v_a &=& \frac{\pi}{\s} \frac {\d }{\d \o^a} S_{ind}= \frac{\pa_-}{\pa_+}\o_a
+\cdots
\ena
we find, as in (1.2),  the Ward identity for $\e$ symmetry
\EQ
\pa_+u = D_1h +\frac{3B}{c}(3\psi_i'q^i +\psi^iq'_i) -\frac{12\s}{c} \o'_av_a
\EN
where
\EQ
D_1 \equiv \pa_-^3 +2u \pa_- +u'
\EN
We have introduced the notation $u' = \pa_-u$, $q_i' = \pa_-q_i$, etc.

For local supersymmetry we will encounter $J_{ab}^{eff}(x) \equiv
\langle J_{ab}(x) \exp S_{int} \rangle $. The OPE for $T(x) J_{ab}(y)$
contains a central term ($TJ$ contains $J$, and $JJ$ contains a central term),
but $Q^iJ_{ab}$ has of course no central term, and also $J_a J_{bc}$ is
without central term (see below).
By direct evaluation one finds
\bea
\underbrace{T(z) :J_aJ_b:(w)} &=& -\s \frac{\d_{ab}}{(z^--w^-)^4} +\e_{abc}
\frac{J^c(w)}{(z^--w^-)^3} \nonumber\\
&+& 2\frac{:J_aJ_b:(w)}{(z^--w^-)^2} +\frac{:J_aJ'_b +J'_aJ_b:(w)}{z^--w^-}
\ena
Since $J_{ab}$ is symmetric in $ab$, one finds
\EQ
\underbrace{T(z)J_{ab}(w)} = -\s \frac{\d_{ab}}{(z^--w^-)^4}+2
\frac{J_{ab}(w)}{(z^--w^-)^2} +\frac{J_{ab}'(w)}{z^--w^-}
\EN
Therefore
we redefine $J_{ab}$ by adding a term with $T$ such that the redefined
operator, denoted by $\L_{ab}$, has no central term in the OPE with
$S_{int}$
\EQ
\L_{ab}(z) = J_{ab}(z) +\frac{2\s}{c} \d_{ab} T(z)
\EN

Another result we shall need is the OPE for $J_a(z)$ and $J_{bc}(w)$.
Before symmetrizing on $bc$ one gets
\bea
\underbrace{J_a(z) :J_bJ_c:(w)} = -\s \e_{abc} \frac{1}{(z^--w^-)^3} +\e_{abd}
\e_{ced} \frac{J^e(w)}{(z^--w^-)^2} \nonumber\\
{}~~~~~~~~
-\s \frac{\d _{ab}J_c(w) +\d_{ac}J_b(w)}{(z^--w^-)^2}
+\frac{\e_{abd}:J_dJ_c:(w)
+\e_{acd}:J_dJ_b:(w)}{z^--w^-}
\ena
Symmetrizing in $b,c$ one finds
\EQ
\underbrace{J_a(z)J_{bc}(w)} =
\frac{(-\s +\frac{1}{2})[\d_{ac}J_b(w)+\d_{ab}J_c(w)] -\d_{bc}J_a(w)}
{(z^--w^-)^2} +\frac{\e_{abd}J_{dc}(w) +\e_{acd}J_{db}(w)}{z^--w^-}
\EN
which is clearly without central charge.

We can therefore compute the terms in $\L_{ab}^{eff}$ which are
quadratic in fields
\bea
&& \L_{ab}^{eff}(\mbox{quadr})(z)   \nonumber \\
&&=\langle (J_{ab}+\frac{2\s}{c}\d_{ab}T)(z)
\frac{1}{2!\pi^2} \int d^2x (hT +\psi^iQ_i +\o^cJ_c) \int d^2y
(hT +\psi^jQ_j +\o^bJ_b) \rangle  \nonumber  \\
 & &
\ena
We will be interested
in the leading terms for $\s \rightarrow \infty$. Now, $T\L$ contains
terms with $T$ and $\L$, see (3.13), but only the former contribute to
$\L_{ab}^{eff}(\mbox{quadr})$ since $\langle \L T \rangle =0$, and they are of
order $\s$. From $Q_i \L_{ab}$ one gets terms with $Q_i$ and $:Q_iJ_a:$,
but their contribution to (3.17)
is of order $B$, i.e., of order $\s$. The leading
terms in $\s$ only come from the OPE of $J^a(x)\L_{ab}(z)$ because it
contains terms of the form $\s J$, see (3.15), and
$\langle J^a(x)J^b(y)\rangle$ is
itself again of order $\s$. In fact
\EQ
\L_{ab}^{eff}(\mbox{quadr}, \s \rightarrow \infty ) = \frac{1}{2\pi^2} \int
\int
d^2xd^2y \langle \underbrace{J_c(x) \L_{ab}(z)} J_d(y) \rangle \o^c(x)\o^d(y)
\EN
because the other contraction $\langle \L_{ab}(z) \underbrace{J_c(x)J_d(y)}
\rangle$ vanishes.
One finds
\EQ
\langle \underbrace{J_c(x)\L_{ab}(z)}J_d(y) \rangle = \s^2\frac{\d_{ac}\d_{bd}+
\d_{bc}\d_{ad}}{
(x^--z^-)^2(y^--z^-)^2}
\EN
Therefore, to second order in $\o^a$,
\EQ
\L_{ab}^{eff}(z) = \s^2 v_a(z)v_b(z) + {\cal O}(\s )
\EN
One can, actually, prove this result to all orders in the fields in
$v_a(z)$ by a Ward identity [11]. Hence, the terms of order $\s^2$ are exact.

With these preparations done, we can return to the supersymmetry Ward identity.
Under $\d \psi^i = \pa_+\eta^i$ one has
\EQ
\d (\exp S_{ind}) = \left \langle -\frac{1}{\pi}\int d^2x \pa_+\eta^i
\underbrace{ Q_i
\left(-\frac{1}{\pi} \right. \int d^2y (hT+\psi^jQ_j +\o^aJ_a}) \left. \right)
 e^{S_{int}} \right\rangle
\EN
The central term in $QQ$ yields the minimal anomaly, while all terms
linear in currents are eliminated by suitable extra terms in the
transformation laws, but a nonlinear term $QQ \sim \L_{ab}$ remains. One
finds then
\EQ
\d (\eta )S_{ind} = \int d^2x \left( \frac{B}{2\pi} \pa_-^2 \eta^i \psi_i
+\frac{2\g}{\pi} \eta^i \psi^j \L_{ij}^{eff} \right)
\EN
under
\bea
\d (\eta )h &=& 2\eta^i\psi_i - \frac{4\g \s}{c} \eta^i \psi_i \nonumber\\
\d (\eta )\psi^i &=& \pa_+\eta^i -h \pa_-\eta^i +\frac{1}{2} \eta^i
\pa_-h - \e_{ija} \eta_j \o_a
\nonumber\\
\d (\eta ) \o^a &=& -\frac{K}{2} \e^{ija}(\eta '_i\psi_j -\eta_i\psi'_j)
\ena
(At this point we already note that we could modify the constant in front
of the last term in (3.22) by adding a nonlinear term $ \d^{nonlinear}\o
\sim \psi \eta v$ in (3.23). In the next section we shall fix this ambiguity
such that the local gauge algebra closes; as a result the sign in front of
the last term in (3.22) changes.)
Extracting the supersymmetry parameters $\eta^i$ one is left with
\bea
(\pa_+-\frac{3}{2}h'-h\pa_-)q^i -\frac{c}{6B}(2-\frac{4\g \s}{c})\psi^i u
+\e^{iaj}\o_aq_j +\e^{ija} (2\psi '_j +\psi_j \pa_-)v_a \nonumber\\
=\pa_-^2\psi^i +\frac{4\g}{B} \psi^j \L_{ij}^{eff}
\ena

Finally, the SO(3) Ward identity is derived without any
complications of nonlinear terms. One finds that
\EQ
\d (\l )S_{ind} = \frac{\s}{\pi} \int d^2x \pa_- \l^a \o_a
\EN
under
\bea
\d (\l ) h &=& 0 \nonumber\\
\d (\l )\psi^i  &=&  \e^{aij} \l_a\psi_j \nonumber\\
\d (\l ) \o^a &=& \pa_+\l^a -h \pa_-\l^a - \e^{abc}\l_b \o_c
\ena
Extracting $\l^a$ we get
\EQ
\pa_+v^a -\e^{aij}\psi_iq_j -h'v^a -h \pa_-v^a+\e^{abc}\o_bv_c = \pa_-\o^a
\EN

For our  purpose we need the terms of leading order in $\s$ in the three Ward
identities. This means that only $S_{ind}^{(0)}$ contributes to the
currents while only the $\s^2$ term in (3.20) survives. One has, in fact,
\bea
(\pa_+ -h \pa_--2h')u -2(3\psi'_iq^i +\psi^iq'_i) +4 \o_a'v^a &=& \pa_-^3h
\nonumber\\
(\pa_+ -h\pa_- - \frac{3}{2}h')q^i -\e^{aij}\o_aq_j -\frac{1}{2}\psi^iu +
\e^{ija}(2\psi_j'+\psi_j \pa_-)v^a -\psi_jv^jv^i &=& \pa_-^2\psi^i
\nonumber\\
(\pa_+-h\pa_--h')v^a +\e^{abc}\o_bv_c - \e^{aij}\psi_iq_j &=& \pa_-\o^a
\nonumber\\
{.}&&{~}
\ena

\sect{Local gauge algebras and the Ward identities for $\s \rightarrow
\infty$.}

In the previous section we obtained the Ward identities for the induced
action $S_{ind}$. For general central charge $\s$, the nonlinearity of
the algebra leads to $\L_{ab}^{eff}$ which is a nonlocal functional
depending on the fields $h$, $\psi^i$, $\o^a$, and the currents $u$,
$q^i$, $v^a$ (the latter are normalized to $\pa_-^3/\pa_+h +\cdots$,
$\pa_-^2/\pa_+\psi^i +\cdots$ and $\pa_-/\pa_+\o^a+\cdots$ by $\s$-independent
rescalings). However, for $\s \rightarrow \infty$, we found local Ward
identities for the leading part of the induced action, $S_{ind}^{(0)}$.
It is these local Ward identities from which we will obtain the one-loop
contributions to the effective action, and it is obviously important to
have a check on their correctness. In addition, we want to establish a
connection between these Ward identities and the gauging of nonlinear
algebras, for which a general formalism was constructed in ref. [12].
In fact, we will see that at the quantum level the anomalies add quantum
corrections to the transformation laws of the currents so that the
nonclosure terms in the classical gauge algebra are eliminated.

Let us begin by emphasizing that the Ward identities are a property
of an induced action, not of particular transformation laws. However, we
can derive them by choosing
certain transformation rules for the gauge fields, and then
varying the induced action under these particular transformation rules.
In our case we determined the
transformation rules for the fields of the form $\d (field) = \pa_+
(parameter) +(field)~\times~(parameter)$ by removing terms in the OPE which
are linear in operators (see section  3).
The
left-over, e.g. the right-hand-sides in (3.7) or (3.22)
is the anomaly. Since the currents are the
Euler-Lagrange equations of the induced action, it is clear that also
terms in the Ward identities which are quadratic in currents can be
(partly or completely) removed from the anomaly by adding terms to the
transformation rules of the gauge fields of the form $\d (field,~extra)
= (field)\times (current) \times (parameter)$. In fact, one needs such
terms if one requires that the currents transform as given by the OPE:
$\d u = \langle \oint \e (x) T(x)dx T(y) \exp S_{int} \rangle$, etc.

It is clear, from the fact that for example $u \sim \langle T \exp S_{int}
\rangle$ and $T$ is holomorphic,
 that the currents transform only into expressions involving
$\pa_-$ derivatives, but no $\pa_+$ derivatives. From this observation
one can immediately read off the transformation laws of the currents from
the Ward identities. Namely, by varying the Ward identities one obtains
expressions of the form $\pa_+(\d ~ current ) + ~more~=0$, and only the
variations of the gauge fields produce further $\pa_+$ derivatives. One
can pull all these $\pa_+$ derivatives in front of the whole term in which they
appear, because the extra terms one produces in this way are of the form
$\pa_+(current)$ which can be rewritten in terms of $\pa_-$ derivatives
by using the Ward identities. As an example consider the term $-h\pa_-u$
in the first Ward identity in (3.28). It
varies into $-\pa_+[\e \pa_-u] +\e \pa_-
[\pa_+u]$ under $\e$ symmetry, and $\pa_+u$ can be replaced by $h
\pa_-u +\cdots$. In this way we find the following transformation rules
for the currents
\bea
\d u &=& \e u' +2\e 'u +6 \eta'_i q^i +2 \eta^iq'_i -4\l'_av^a
+\pa_-^3\e \nonumber\\
\d q^i &=& \e q'^i +\frac{3}{2} \e 'q^i - \e^{iaj}\l_aq_j
 +\frac{1}{2}\eta^iu-\e^{ija}
(2 \eta '_j +\eta_j \pa_-)v_a
+ \eta_j v^jv^i +\pa_-^2\eta^i \nonumber\\
\d v^a &=& \e v'^a +\e 'v^a -\e^{abc}\l_bv_c +\e^{aij}\eta_iq_j +\pa_-\l^a
\ena
Note that these results hold only for $\s \rightarrow \infty$ because we
used the $\s \rightarrow \infty$ Ward identities, but we could also obtain
the results for finite $\s$ by using the same ideas. Note also that in
this derivation we only used the leading term in the gauge field
transformation laws (the $\pa_+ (parameter)$ part) and the possibility
of extra terms in the gauge field transformation laws involving currents
is still completely left open. For further use we draw the reader's
attention to the one nonminimal term in $\d q^i$, and to the minimal
anomalies $\pa_-^3 \e$, $\pa_-^2 \eta^i$, and $\pa_-\l^a$.

We have, in  fact, used in obtaining (4.1) that the Ward identities in (3.28)
can be written in
terms of {\em supercovariant} derivatives as
\EQ
(D_+j)_A = \eta_{AB}\pa_-^{4-B}\phi^B ~~~~,~~~A,B=1,2,3
\EN
where $j_1 =u$, $j_2 = q_i$, $j_3 = v_a$, and $\phi^1 =h$, $\phi^2 =
\psi^i$, $\phi^3 = \o^a$, while $\eta_{AB} = \d_{AB}$.
By supercovariance of $D_+$ we mean that the variation of $D_+j$ is
independent of $\pa_+ \xi^A$ if $\xi^A$ are the local parameters.
This allows us to determine $\d j_A$.

We could now deduce the transformation rules of the gauge fields by
requiring that they, together with the current laws given above, leave
the Ward identities invariant. However, there is a simpler, more general
and more elegant method, and that is to note that given a nonlinear algebra
of the form
\EQ
{~} [\hat{T}_A, \hat{T}_B] =\hat{T}_Cf^C_{~AB}
+\hat{T}_D\hat{T}_CV^{CD}_{~~~AB}
\EN
one can gauge it. One can then {\em derive} the following transformation rules
of the gauge fields $h_{\m}^{~A}$ and ``auxiliary fields'' $T_A$
\bea
\d h_{\m}^{~A} &=& \pa_{\m}\e^A +f^A_{~BC}h_{\m}^{~C} \e^B +T_DV_{~~BC}^{DA}
h_{\m}^{~C}\e^B
\nonumber\\
\d T_A &=& T_C(f^C_{~AB}+\frac{1}{2}T_DV^{DC}_{~~AB})\e^B
\ena
and find the following results [12]:

\noindent (i) the gauge  commutator on gauge fields closes up to a
covariant derivative
\EQ
[\d (\e_1), \d (\e_2)] h_{\m}^{~A} = \d (\e_3)h_{\m}^{~A} -D_{\m}T_D
V^{DA}_{~~BC} \e_1^{~C}\e_2^{~B}
\EN
where the covariant derivative of $T_A$ is given by
\EQ
D_{\m}T_A = \pa_{\m}T_A -
 T_Cf^C_{~AB}h_{\m}^{~B}-\frac{1}{2}T_CT_DV^{DC}_{~~AB}h_{\m}^{~B}
\EN
Note the factor $\frac{1}{2}$ in $\d T_A$ and $D_{\m}T_A$.

\noindent (ii) The covariant derivatives are really covariant: they
transform in the coadjoint representation, defined by
\EQ
\d D_{\m}T_A = D_{\m}T_C \tilde{f}^C_{~AB}\e^B
\EN
where $\tilde{f}^C_{~AB}$ are field-dependent structure  constants
\EQ
\tilde{f}^C_{~AB} = f^C_{~AB} +T_DV^{DC}_{~~AB}
\EN

\noindent (iii) The gauge commutator on the auxiliary fields $T_A$ closes
\bea
&&[\d (\e_1), \d (\e_2)]T_A = \d (\e_3)T_A \nonumber\\
&&\e_3^{~C} = \tilde{f}^C_{~AB}\e_1^{~B}\e_2^{~A}
\ena

\noindent (iv) curvatures are defined by
\EQ
[D_{\m},D_{\n}]T_A = -T_C(f^C_{~AB}+\frac{1}{2}T_DV^{DC}_{~~AB})R_{\m\n}^{~~B}
\EN
and transform as follows
\EQ
\d R_{\m \n}^{~~A} = \tilde{f}^A_{~BC}R_{\m \n}^{~~C}\e^B
+D_{\m}T_DV^{DA}_{~~BC}h_{\n}^{~C}\e^B - \m \leftrightarrow \n
\EN
They satisfy the Bianchi identities
\EQ
D_{[\m}R_{\n \r ]}^{~~~A} =0
\EN

We shall now make contact with our Ward identities by the following
observations:

\noindent (i) the auxiliary fields $T_A$ are identified with the
currents $j_A$ in the Ward identities.

\noindent (ii) the formalism for gauging a nonlinear algebra holds for
classical algebras. The Ward identities, however, contain a minimal
anomaly term ( the term $\pa^{4-B}\phi^B$ in eq. (4.2)), and this term
leads to a quantum correction $\d_qT_A$.

\noindent (iii) the covariant derivatives $D_{\m}T_A$ are the Ward
identities except for the minimal anomaly.

In fact, using our rule of
extracting $\pa_+$ derivatives from the variations of the Ward
identities, we already found the extra quantum terms: $\pa_-^3\e$ in $\d
u$, $\pa_-^2\eta^i$ in $\d q^i$, and $\pa_-\l^a$ in $\d v^a$, see (4.1). Since
these terms are field-independent, the gauge commutator on the currents
$j_A$ must still close (it indeed does, see below). However, also on the
gauge fields the gauge commutator now closes because the quantum
term in $\d T_A$ precisely cancels the covariant derivative in eq. (4.5).
In other words, the minimal anomaly completes the Ward identity so that
the gauge commutator closes on fields and currents! Finally, since there
is a nonlinear term in $\d T_A$ (namely the term $\d q_i \sim \eta^j
v_jv_i$), we predict one nonlinear term in the gauge field laws as well
\EQ
\d^{nonlinear} \o^a = (\psi^a\eta^b +\psi^b \eta^a)v_b
\EN
With this, we have checked that the gauge algebra on fields and currents
closes. The commutator of two supersymmetries is, as always, the most
interesting,
\bea
&&[\d (\eta_1), \d (\eta_2)] = \d (\hat{\e}=2\eta_2^i \eta_1^i) +\d
(\hat{\l} ^a) \nonumber\\
&&\hat{\l}^a = \e^{aij}(\eta_{2i}\eta '_{1j} - \eta_{1i}\eta '_{2j}) +
(\eta_1^a \eta_2^b-\eta_2^a\eta_1^b)v_b
\ena
The $v$-term in $\hat{\l}^a$ is due to the structure constants $V$ in (4.3).
 The rest of the gauge algebra is as expected
 \bea
 {~} [\d (\eta ),\d (\e )] &=& \d (\hat{\eta}_i = \e \eta '_i
-\frac{1}{2}\e ' \eta_i ) \nonumber\\
{~}[ \d (\eta ), \d (\l )] &=& \d (\hat{\eta}^i = \e^{ibk}\l_b\eta_k )
\nonumber\\
{~}[ \d (\e_1), \d (\e_2)] &=& \d ( \hat{\e}= \e_2\e '_1 -\e_1\e '_2)
\nonumber\\
{~}[ \d (\l ), \d (\e )] &=& \d (\hat{\l}_a = \e \l '_a ) \nonumber\\
{~}[ \d (\l_1), \d (\l_2)] &=& \d (\hat{\l}_a =2 \e_{abc}\l_1^b\l_2^c)
\ena

The extra variation (4.13) gives an extra contribution to the anomaly in
(3.22) which is of the same form as the last term. In fact, the net result of
including this variation is to change the sign of the nonlinear anomaly in
(3.22).

\sect{The SO(N) theories from constrained WZWN models.}

The SO(N) models are based on nonlinear superconformal algebras
which contain the same set of generators as one encounters in the linear
superalgebras Osp(N$|$2). It points to a close relation between the nonlinear
SO(N) theories and WZWN models based on Osp(N$|$2). In fact, it has been shown
in ref. [3] that Polyakov gravity or $W_3$ gravity can be obtained from
Sl(2,R) or Sl(3,R) induced Yang-Mills theory by imposing constraints on the
{\em currents}. This suggests imposing constraints on the currents $u \equiv
u^aT_a$ of induced Yang-Mills theory with $T_a$ the generators of Osp(N$|$2)
(see also ref. [13]).
The constraints for Osp(1$|$2) which produce (1,0) supergravity were already
given in [3].
Our aim is to obtain the Ward identities (3.28) from the Ward identities
of Yang-Mills by imposing constraints on the Yang-Mills currents and by
making suitable identifications between the Yang-Mills gauge fields
and currents and those of SO(N) supergravity. We will give the details for the
case N=3.

Since we need all SO(3) connections $\o^a$ in the
nonlinear theory, we cannot put constraints in the SO(3) sector of $u$.
Hence, if the SO(3) models can be obtained at all by imposing constraints on
Osp(3$|$2), these constraints {\em must} be of the following form
\bea
u^aT_a =
\left( \begin{array}{ccccc}
0&u^{\neq}&u^{+1}&u^{+2}&u^{+3} \\
1&0&0&0&0\\
0&-u^{+1}&u^{11}&u^{12}&u^{13}\\
0&-u^{+2}&u^{21}&u^{22}&u^{23}\\
0&-u^{+3}&u^{31}&u^{32}&u^{33}
\end{array} \right)
\ena
with $u^{ij}=\e^{ija}u^a$. The Yang-Mills fields $A=A^aT_a$ are not
a priori constrained
\bea
A^aT_a =
 \left( \begin{array}{ccccc}
A^0&A^{\neq}&A^{+1}&A^{+2}&A^{+3}\\
A^{=}&-A^0&A^{-1}&A^{-2}&A^{-3}\\
A^{-1}&-A^{+1}&A^{11}&A^{12}&A^{13}\\
A^{-2}&-A^{+2}&A^{21}&A^{22}&A^{23}\\
A^{-3}&-A^{+3}&A^{31}&A^{32}&A^{33}
\end{array} \right)
\ena
with $A^{ij}=\e^{ija}A^a$.
 However, substituting these expressions for $u$ and $A$ into the Ward
identity of the induced Osp(3$|$2) Yang-Mills theory, cf. (1.2)
\EQ
\pa_+u=\pa_-A +[A,u]
\EN
all components of $A$ are expressed in terms of the following fields:
$A^=$, $A^{-i}$, $A^a$. For example, from $\pa_+u^==0$ one finds that
\EQ
A^0=\frac{1}{2}(A^=)'
\EN
while from $\pa_+u^0=0$ one obtains
\EQ
A^{\neq}= -\frac{1}{2}(A^=)'' +u^{\neq}A^= +u^{+i}A^{-i}
\EN
The constraint $u^{-i}=0$ leads to
\EQ
A^{+i}=(A^{-i})'+u^{+i}A^=+ \e^{ijk}u^jA^{-k}
\EN
Substituting these relations into the remaining Ward identities for $u^{\neq}$,
$u^{+i}$ and $u^a$ one finds the following results
\bea
\pa_+u^{\neq} &=& -\frac{1}{2}(A^=)'''+(u^{\neq}A^= +u^{+i}A^{-i})'
+(A^=)'u^{\neq} \nonumber\\
&&-2[(A^{-i})'+u^{+i}A^= +\e^{ijk}u^jA^{-k}]u^{+i} \nonumber\\
\pa_+u^{+i} &=& (A^{-i})''+(u^{+i}A^=+\e^{ijk}u^jA^{-k})' -A^{-i}u^{\neq} -
\e^{ijk}A^ju^{+k} \nonumber\\
&&+\frac{1}{2}(A^=)'u^{+i} -\e^{ijk}[(A^{-j})'+u^{+j}A^= ]u^k
+u^iu^kA^{-k} -A^{-i}u^ku^k
\nonumber\\
\pa_+u^{ij}&=& (A^{ij})' +(A^{-i}u^{+j}-A^{-j}u^{+i}) +\e^{ikl}\e^{lmj}
(A^ku^m-A^mu^k)
\ena

It is clear that one can choose the scales such that
$A^=$ is equal to $h$, since there is only one term with
$h'''$. Similarly, we can identify $A^{-i}$ with $\psi^i$, but for $A^a$ there
is an ambiguity allowed by dimensional arguments and index structure: $A^a=
\o^a +\a h v^a$. For the currents, $u^{+i}=q^i$ and $u^a=v^a$ are the only
possibilities, but $u^{\neq}= -\frac{1}{2}u +\b v^av_a$ is possible.
Substituting these identifications into (5.7), we find that with
\bea
u^{\neq} &=& -\frac{1}{2}u -v^av^a
\nonumber\\
A^a &=& \o^a +hv^a
\ena
one reproduces indeed the Ward identities in (3.28). In particular,
 the nonlinear
term in the $\pa_+q^i$ Ward identity is due to the term $\e^{ijk}A^{+j}u^k$
after substituting $A^{+j}= \e^{jpq}u^pA^{-q}+ \cdots$. The reason there are
no further nonlinear terms in the Ward identities is that terms with
$q^iq^i$ or $\e^{ibc}v^bv^c$ are produced, which obviously cancel, while the
redefinitions in (5.8) remove some nonlinear terms.

One can now substitute the constraints on the currents into the transformation
 laws of the currents. The latter follow from $u=\del_{-}gg^{-1}$ and read
$\d u=\del_{-}\h +[\h ,u]$. In this way one finds that only the parameters
$\h^{=}$,$ \h^{-i}$ and $\h^a$ are unconstrained, and furthermore one
recovers (4.1), including the minimal anomaly terms. Repeating the analysis
for the gauge fields, one finds from $\d A=\pa_{+}\h +[\h ,A]$ and
the relations between the parameters which we mentioned above, that the gauge
fields transform as given in section 3 and (4.13).

It is also possible to obtain the $\s \rightarrow \infty$ part of the induced
action, $S_{ind}^{(0)}(h, \psi , \o )$, in closed form from the WZWN action by
using the identifications between the currents. Before doing so, we must first
discuss a subtlety concerning the indices of the currents. In
section 3 all currents have lower indices because the Ward identities follow
from $\d \phi^A j_A= \phi^A O_{AB}\pa \xi^B$, where $j_A \sim (\d /\d
\phi^A )
S$.
Even in the term nonlinear in currents we identified $\frac{\pa_-}{\pa_+}
\o^a$ with the current $\d S/\d \o^a =j_a$. On the other hand, in this section
all currents have upper indices. Hence, the identifications we made were
really as follows:
\bea
A^= =h ~~~;~~~A^{-i} = \psi^i ~~~;~~~A^a = \o^a +h \d^{ab}v_b \nonumber\\
u^{+i} = \d^{ij}q_j~~~;~~~u^a=\d^{ab}v_b ~~~;~~~u^{\neq}=-\frac{1}{2}u -
\d^{ab}v_av_b
\ena
The numerical constants $\d^{ab}$ and $\d^{ij}$ were introduced by the OPE
given in section 2.

Consider now the relation $u^{\neq}=-\frac{1}{2}u-\d^{ab}v_av_b$.
We can substitute the definition of the currents
using $u^{\neq}=g^{\neq , =}u_=$. Recall that
\bea
u_==2\pi \frac{\d S_{ind}^{(0)}(A)}{\d A^=} ~~~,~~~ u=-4\pi \frac{ \d
S_{ind}^{(0)}(h,\psi ,\o )}{\d h} \nonumber\\
q_i=-\pi \frac{\d}{\d \psi^i}S_{ind}^{(0)}(h, \psi , \o )
{}~~~,~~~v_a= \pi \frac{\d S_{ind}^{(0)}(h,\psi , \o )}{\d \o^a }
\ena
We obtain then
\EQ
\left. \left( \frac{\d}{\d h}-\frac{1}{2}\d^{ab}v_a \frac{\d}{\d \o^b} \right)
S_{ind}^{(0)}(h, \psi ,\o ) = g^{\neq,=} \left( \frac{\d S_{ind}^{(0)}(A)}
{\d A^=} \right) \right|
\EN
where the bar on the right hand  side indicates that one should express all
fields $A^a$ in terms of the unconstrained final fields $H=\{ A^==h,
A^{-i}=\psi^i,A^a=\o^a+h\d^{ab}v_b\}$ {\em after}
one has performed the variation
with respect to $A^=$. If we invert the order of variation and imposing
constraints we get extra terms, due to the chain rule. Namely, we find the
following equality
\bea
\left. \left( \frac{\d S_{ind}^{(0)}[A]}{\d A^=}\right) \right|
=&& \frac{\d S_{ind}^{(0)}
[A(H)]}{\d h} \\
&&-  \left. \left( \frac{\d S_{ind}^{(0)}[A]}{\d A^{\neq}}\right) \right|
 \frac{\d A^{\neq}
(H)}{\d h}
-\left. \left(\frac{\d S_{ind}^{(0)}[A]}{\d A^a} \right) \right|
\frac{\d A^a(H)}{\d h} \nonumber
\ena
There are no extra terms coming from varying $A^0$ or $A^{+i}$ with respect
to $h$ because $u^0$ and $u^{-i}$ vanish due to the constraints. Moreover,
using that $A^a = \o^a +h\d^{ab}v_b$,
$u^{=}=1$, and $u_a=g_{ab}v_b$, we see that
\bea
\left( \frac{\d}{\d h} -\frac{1}{2} \d^{ab}v_a \frac{\d}{\d \o^b} \right)
S^{(0)}_{ind}(h, \psi , \o ) =g^{\neq ,=} \frac{\d}{\d h} S_{ind}^{(0)}
(A[H]) \nonumber\\
{}~~~~~~~~~~~~~~
 - \frac{1}{2 \pi} \int d^2x \frac{\d A^{\neq}}{\d h} - g^{\neq , =}g_{ab}
\frac{1}{2 \pi} \int d^2x u^b \frac{\d}{\d h} (hv_a)
\ena
Since $g^{\neq ,=}g_{ab}= +2 \d_{ab}$ ( the positive sign is due to the
supertrace), both sides can be written as a $\d / (\d h) $ derivative
(the second term on the left hand side cancels against the extra piece
from the last term on the right hand side), and
we conclude that
\EQ
S_{ind}^{(0)}= g^{\neq , =} S_{ind}^{(0)}(A[H]) -\frac{1}{2\pi} \int d^2x
(A^{\neq} +h \d^{ab}v_av_b )
\EN
The total derivative $-\frac{1}{2} (A^=)''$ in $A^{\neq}$, eq. (5.5),
 can be dropped, and
replacing $v_a$ and $v_b$ by
$u^a$ and $u^b$, we obtain the induced
action of (N,0) supergravity in terms of the action and currents of the
corresponding WZWN model.

As a check, we may repeat the derivation by using $u^{+i} = \d^{ij}q_j$ or
$u^a = \d^{ab}v_b$. In the former case one finds
\bea
\frac{\d}{\d \psi^i} S_{ind}^{(0)}(h, \psi , \o )=
&&\left(\frac{-2g^{+i,-i}}{g^{\neq , =}} \right)
\left [ g^{\neq , =}\frac{\d}{\d \psi^i} S_{ind}^{(0)} (A[H]) \right.
\\
&&\left.  -\frac{1}{2\pi} \int d^2x \frac{\d}{\d \psi^i} A^{\neq}
-\frac{1}{2\pi}
(g^{\neq , =}g_{ab}) \int d^2x v_b \frac{\d}{\d \psi^i} (hv_a) \right]
\nonumber
\ena
which agrees with (5.14) since $g^{+i,-i}/g^{\neq , =} =- \frac{1}{2}$
(using the fact that
$str T_{+i}T_{-i} =2= -str T_{-i}T_{+i}$ and $str T_{\neq}T_= =1$).
In the latter case we
may note that differentiation of $S_{ind}^{(0)}$ with respect to $A^b$ and
then imposing constraints, differs from $(\d S_{ind}^{(0)} / \d A^c )(\d A^c
/ \d \o ^b)$ by a term proportional to $u_c \frac{\d}{\d \o^b}hv_c =
\frac{1}{2} \frac{\d}{\d \o^b} (h v_cv_d g_{cd})$. We get then
\bea
S_{ind}^{(0)}(h, \psi , \o ) =&&\left( \frac{2g^{aa}}{g^{\neq , =}}\right)
g^{\neq , =} S_{ind}^{(0)}(A[H]) \nonumber \\
&&-\frac{1}{\pi} \left( \frac{g^{aa}}{g^{\neq , =}} \right) \int d^2x A^{\neq}
-
g^{aa}g_{cd} \frac{1}{2\pi} \int d^2x h v_cv_d
\ena
(with no summation over $a$ in $g_{aa}$), which again agrees with (5.14)
if we use that
$g^{aa}/g^{\neq , =} =\frac{1}{2}$.

Summarizing,
the $\s \rightarrow \infty$ part of
the action for induced supergravity is equal to the WZWN action in which the
constraints have been inserted, plus correction terms which are due to the
noncommutativity of varying and imposing constraints, and which depend on the
currents of the WZWN model in which the constraints have been substituted.

The method of putting constraints on the currents of a WZWN model and
obtaining Ward identities for a model based on a nonlinear superalgebra can be
pulled back to the algebraic level. We expect that the U(n) nonlinear
superalgebra given in refs. [6,7] can be obtained in the same manner from
the corresponding linear SU(n$|$2) superalgebras, but note that for these
models no BRST charge seems to exist [10].

\sect{Computation of one-loop contributions to the effective action.}

As explained in the introduction and in ref. [5],
we obtain the one-loop contributions to the
effective action by taking the determinants of the matrices $M$ and $N$,
obtained by differentiating the $\s \rightarrow \infty$ Ward identities with
respect to the gauge fields. The matrix $M$ is the covariant derivative
\bea
M=
 \left( \begin{array}{ccc}
\nabla^{2}_{+}&4\o '_b &-6\psi '_j-2\psi_j\pa_-  \\
{}~&~&\\
0& \nabla_{+}^{1}\d^a_{~b} +\e^a_{~cb}\o^c &  \e^a_{~jk}\psi^{k}\\
{}~&~&\\
-\frac{1}{2}\psi^i &~~\e_b^{~ik} (2\psi '_k+\psi_k\pa_-) -\d^i_{~b}\psi^kv_k
-v^i\psi_b &\nabla^{\frac{3}{2}}_{+}\d^i_{~j}-\e^i_{~ja}\o^a
\end{array} \right) & & \nonumber \\
 & &
\ena
where $\nabla^{j}_{+} =\del_{+}-h\del_{-}-jh'$.

The matrix $N^a_{~b}$ is its dual in the sense of the introduction
\bea
N=
\left( \begin{array}{ccc}
\pa_-^3+2u\pa_-+u' & -4v_b\pa_- & -6q_j\pa_--2q'_j \\
{}~&~&\\
v^a\pa_- +(v^a)' & \pa_-\d^a_{~b} -\e^a_{~bc}v^c & -\e^a_{~jk}q^k \\
{}~&~&\\
(q^i)' +\frac{3}{2}q^i\pa_- & \e_b^{~ik}  q_k & (\pa_-^2+\frac{u}{2})\d^i_{~j}
+v^iv_j -\e^{i}_{~ja}(2v^a\pa_- +(v^a)')
\end{array} \right) & &  \nonumber \\
 & &
\ena

The results for the $hh$, $\psi^i\psi^i$  and $\o^a\o^a$ self-energies are
given below in (6.3). In each case one gets one contribution
from  $\frac{1}{2} \ln
{\rm sdet} M$ by evaluating a self-energy graph with two vertices, each linear
in the
external field and bilinear in internal ghosts, and another contribution from
$-\frac{1}{2} \ln {\rm sdet} N$, by evaluating
the same graphs, but now with vertices linear in currents. For the currents
we take the leading terms (which are linear in fields). In the loop one
finds the anticommuting antighosts-ghosts $b_1 , c^1 ,b_{2a} ,c^{2a}$ and
$B_1 , C^1$, $B_{2a} , C^{2a}$, corresponding to the bosonic fields
and  the
commuting pair $b_{3i},c^{3i}$ and $B_{3i} , C^{3i}$ corresponding to the
fermions. (We have ordered the rows and
columns of $M$ and $N$ such that the fermions are in the last row and
column.) We obtain
\bea
\begin{array}{cccc}
{}~& \langle hh \rangle &\langle \o^a \o_a \rangle & \langle \psi^i \psi_i
\rangle \\
{}~&~&~&~\\
\frac{1}{2}\ln {\rm sdet} M: & -\frac{1}{48\pi}h\frac{\pa_-^3}{\pa_+}h & 0 &
-\frac{1}{4\pi} \psi^i\frac{\pa_-^2}{\pa_+}\psi_i \\
{}~&~&~&\\
-\frac{1}{2} \ln {\rm sdet} N:& -\frac{1}{8\pi} h \frac{\pa_-^3}{\pa_+}h &
\frac{1}{2\pi} \o^a \frac{\pa_-}{\pa_+}\o_a &- \frac{1}{2\pi} \psi^i
\frac{\pa_-^2}{\pa_+}\psi_i
\end{array}
\ena

We comment briefly on these results. For the graviton self-energy $\langle hh
\rangle$, the numerical factor in the $M$ contribution is
$\frac{1}{4}(j^2-j+\frac{1}{6})$, summed
over $j=2, \frac{3}{2},1$, and is a factor $-1/26$ smaller than the
pure gravity case. (The sign is negative because the supersymmetry ghosts
win.)
The $N$ contribution is $-\frac{1}{24}j(2j+1)(2j+2)$ for spin $j+1$.
Hence it only comes from coordinate and supersymmetry ghosts, with
contributions $-\frac{1}{2}$  and $3\times (\frac{1}{8})$ respectively,
giving a
result which is a factor 1/4 smaller than in pure gravity. The reason the
Yang-Mills ghosts do not contribute to $N$ is that on dimensional
grounds a coupling to the
source $u$ is not possible.
For the Yang-Mills
self-energies, there is no contribution from $M$, because the 3 pure
supersymmetry ghost loops cancel the 3 pure Yang-Mills ghost loops, while no
mixed loops with
off-diagonal vertices can be  constructed due to the triangular
structure of the $\o^a$ couplings in $M$. For $N$, the coordinate ghosts yield
a vanishing contribution, while the Yang-Mills ghost loop yields a
factor $-\frac{1}{2}$, and the supersymmetry
ghost loop a factor +1. Finally, the
gravitino selfenergy $\langle \psi^i\psi_i \rangle$ receives one $M$
contribution $\frac{5}{4}$ from the mixed $b_3,c_3,b_1,c_1$ loop, and
another $M$ contribution $-\frac{3}{2}$ from the mixed $b_3,c_3,b_2,c_2$
loop. The $N$ contributions come from similar loops and are given by
$-\frac{3}{2}$ and $+1$ respectively.

We can, in fact, easily find the one-loop contributions to the effective
action with any number of $h$ fields
\EQ
S_{eff}^{1-loop} (all ~h) = -
\frac{1}{26}[ -13 S_{ind}^{(0)}(h)] +\frac{1}{4}[ 12 \bar{S}_{ind}^{(0)}(u)]
\EN
where
\EQ
S_{ind}^{(0)}(h) = -\frac{1}{24\pi}\int h \frac{\pa_-^3}{\pa_+}h +\cdots
\EN
is the Polyakov action in (1.8) and (1.9), and where
\EQ
\bar{S}_{ind}^{(0)}(u) = -\frac{1}{24\pi}\int u \frac{\pa_+^3}{\pa_-}u +\cdots
\EN
is its dual in the sense of section 1. They are related by a Legendre
transformation
\EQ
S_{ind}^{(0)}(h) +
\bar{S}_{ind}^{(0)}(u(h)) = -\frac{1}{12\pi} \int d^2x h u(h)
\EN
Hence, in the $h$ sector we find
\EQ
S_{eff}^{1-loop} (all ~h) = -\frac{5}{2} S_{ind}^{(0)}(h) -\frac{1}{4\pi}
\int d^2x hu
\EN

The complete one-loop effective action is obtained
by an overall rescaling of the central charge $\s $ by a factor $Z_{\s}$ and
rescalings of the gauge fields:
\EQ
\s S_{ind}^{(0)}(h, \psi , \o ) +S_{eff}^{1-loop} (h, \psi , \o ) =
Z_{\s} \s  S_{ind}^{(0)}( Z_hh, Z_{\psi}\psi , Z_{\o} \o )
\EN
{}From the result in (6.8), plus those for the two-point functions,
we deduce that at the one-loop order
\bea
&&Z_{\s}= 1-\frac{5}{2\s} ~~~~,~~~~Z_h=1+\frac{11}{6\s} \nonumber\\
&&Z_{\o}=1+\frac{7}{4\s} ~~~~,~~~~Z_{\psi}=1+\frac{2}{\s}
\ena

We have repeated the calculations performed in this section for the N=1
and N=2 cases. In the N=1 case one simply drops the $\o^a$ and $v^a$ fields
and the index $i$ on the $\psi$ field from the $M$ and $N$ matrices, while
in the N=2 case one has a single $\o$, $v$ pair, with $\e^{aij} \rightarrow
\e^{ij}$, $i,j=1,2$. In both cases the results are (rigidly)
supersymmetric, i.e.
all the fields are rescaled by a common wave-function renormalization factor,
a feature which is lost in the N=3 case. To understand this, we are at
present making a general study of rigid symmetries in models such as these.

\sect{Conclusions.}

We have shown that by imposing constraints on the currents of a WZWN model
based on the linear superalgebra Osp(N$|$2) one obtains the Ward identities for
the induced action based on the nonlinear SO(N) superconformal algebra of
Knizhnik [6] and Bershadsky [7] in the limit of large central charge.
One can also find the
induced action in closed form; it is nonlocal,
and contains a one-component graviton, $N$ chiral gravitinos, and
$\frac{1}{2} N(N-1)$ chiral Yang-Mills fields, and is a (N,0) supergravity
theory in d=2 dimensions. We also computed the one-loop corrections to the
self-energies of these gauge fields, and to the Green's functions with
$n$ external gravitons. They are finite, and the effective action is obtained
from the induced action by a rescaling of the fields and central charge.
These results are quite similar to
those for $W_3$ gravity performed in ref. [3,4,5].

We emphasize that the simple relation between the effective and induced
actions seems to be a consequence of working in chiral gauge, and is not so
apparent when one imposes constraints on WZWN models which lead to Toda-like
actions [14].
In the absence of an all-order
rigorous proof of the conjectured relation between
the induced action and the effective action,  and its
all-loop finiteness, it would be useful
to calculate the two-loop corrections for our model or
one of the models mentioned in the
introduction. The Feynman rules and regularization of higher loops in these
nonlocal chiral field theories has been discussed in ref. [15], but the usual
aspects of local quantum field theory do not apply, so that many issues
remain to be settled (see [15] for a detailed discussion).

We have shown that the Ward identities for an
induced action in the limit
that the central charge tends to infinity are of the form "covariant
derivative of current = minimal anomaly". This led to an interesting
extension of the classical gauging of algebras to the quantum level. Namely,
in the classical gauging of nonlinear algebras, there appears for each gauge
field a corresponding auxiliary field [12], but the local gauge algebra only
 closes
on these auxiliary fields, whereas on the gauge fields one finds extra terms
proportional to the covariant derivatives of the auxiliary fields. By
identifying these auxiliary fields with the currents and
adding the minimal anomalies as quantum corrections to the classical
transformation rules of the currents, one obtains an extra term in the gauge
commutator on the gauge fields, which cancels the covariant derivative of the
current, so that
the {\em classical nonclosure} turns into {\em quantum closure}. The reason
for this remarkable cancellation is quite general (it was already found
to hold for $W_3$ gravity [3]): the covariant derivative of
the auxiliary fields is the Ward identity minus the minimal anomaly, since the
auxiliary fields are nothing else but the currents. This quantum improvement
of a classical imperfect theory suggests further interesting possibilities to
which we hope to return.

Because of the presence of the factor $1/2$ in the covariant derivative of
the currents $D_{\m}T_A$ in (4.6), but absence of a corresponding factor
$1/2$ in the gauge field variations  $\d h_{\m}^A$ in (4.4), the response of
the induced action under {\em these} variations $\d h_{\m}^A$ is the minimal
anomaly {\em minus} the product of the nonlinear terms in the Ward identity
times the gauge parameter. As observed in [3], if one would halve the
nonlinear terms in  $\d h_{\m}^A$  one would completely cancel the nonlinear
terms in the anomaly, but as we have explained here, only the $\d h_{\m}^A$
in (4.4) which came from the gauging of nonlinear algebras will lead to a
closed gauge algebra.

A final comment concerns the integrability conditions of the differentiated
Ward identities. Since $\d u_{-,a}/\d A_+^b = M^{-1}_{ac}N^c_{~b}$ in (1.16)
is symmetric in $a,b$ by virtue of the definition of the currents $u_{-,a}$,
the integrability conditions read $M^{-1}N = N^TM^{-1,T}$, where the
derivatives in $N^T$ and $M^{-1,T}$ act to the left. Partially integrating
them, one obtains operators $N^t$ and $M^{-1,t}$ in which all derivatives act
to the right. Hence one obtains the conditions
\EQ
MN^t-NM^t=0
\EN
For induced Yang-Mills theory, $M=D_+(A)=-M^t$ and $N=D_-(u)=-N^t$, so that
one obtains $ [D_+(A),D_-(u)]=R_{+,-}=0$, which is the well-known
parallelizability of the WZWN model. For Polyakov gravity $N=-N^t=D_1
=\pa_-^3+2u\pa_-+u'$, but $M= \nabla_+^{2}$ and $M^t = - \nabla _+^{-1}$,
where $\nabla_+^j =\pa_+-h\pa_--jh'$. The integrability conditions now yield
(with $D_+u \equiv \nabla _+^2u$)
\EQ
2(D_+u-h''')\pa_- +(D_+u-h''')'=0
\EN
which is indeed satisfied as long as the Ward identity $D_+u-h'''=0$ is
satisfied. More generally, for nonlinear (super)algebras, the consistency
conditions are obtained by replacing each current $T_B$ in the matrix $N$ by
the corresponding Ward identity. Hence, using (4.6)
\EQ
[D_+T_C- (Anomaly)_{+,C}] \tilde{f}^C_{~AB}=0
\EN
where $\tilde{f}^C_{~AB}=f^C_{~AB} +T_DV^{DC}_{~~AB}$ and $(Anomaly)_{+,C}$
is the
minimal anomaly. In practice (7.1) yields a good check on the Ward
identities. One may also view it as a {\em quantum} curvature for chiral
gauge theories, which replaces the classical curvature $R_{\m \n}^{A} =
\pa_{\m}h_{\n}^A - \pa_{\n}h_{\m}^A +\tilde{f}^A_{~BC}h_{\m}^Ch_{\n}^B$ of
nonchiral gauge theories proposed in [12].

\end{document}